\documentclass[12pt ,floatfix,tightenlines]{revtex4}
\usepackage{graphicx,graphics}
\usepackage{amsmath,epsf,subfigure}
\usepackage{color}
\usepackage{amsfonts,amsmath,graphics}
\usepackage{bm,subfigure}
\usepackage{graphicx,epsf}
\usepackage{amssymb}
\usepackage{afterpage}
\usepackage{float}
\usepackage{chemmacros}
\usepackage{tikz-cd}
\usepackage{MnSymbol}
\usepackage{amsmath}
\usepackage{comment}
\newcommand{\obj}{species}

\newcommand{\beq}{\begin{equation}}
\newcommand{\eeq}{\end{equation}}
\newcommand{\bea}{\begin{eqnarray}}
\newcommand{\eea}{\end{eqnarray}}

\renewcommand{\k}{\kappa}

\newcommand{\ys}{\textcolor{black}}

\usepackage{soul}

\begin{document}

\title{On the exclusion of exponential autocatalysts\\
by sub-exponential autocatalysts}

\author{Yann Sakref}
\affiliation{Gulliver, CNRS, ESPCI Paris, Universit\'e PSL, 75005 Paris, France}

\author{Olivier Rivoire}
\email{olivier.rivoire@espci.fr}
\affiliation{Gulliver, CNRS, ESPCI Paris, Universit\'e PSL, 75005 Paris, France}

\begin{abstract}
Selection among autocatalytic \obj~fundamentally depends on their growth law: exponential \obj, whose number of copies grows exponentially, are mutually exclusive, while sub-exponential ones, whose number of copies grows polynomially, can coexist. Here we consider competitions between autocatalytic \obj~with different growth laws and make the simple yet counterintuitive observation that sub-exponential \obj~can exclude exponential ones while the reverse is, in principle, impossible. This observation has implications for scenarios pertaining to the emergence of natural selection.
\end{abstract}

\maketitle

\section{Introduction}

Autocatalysts are molecules that catalyze their own formation, leading to an auto-amplification process: the presence of more autocatalysts of a \obj~leads to a decrease in the time required to produce additional autocatalysts of the same \obj~\cite{SemenovAC}. This auto-amplification can cause the number of autocatalysts of a \obj~to grow at a rate proportional to their concentration, a relationship mathematically described by $dA/dt = kA$, where $A$ denotes the concentration of the autocatalytic \obj~and $k$ its replication rate. This model results in an exponential growth dynamics, $A(t) \sim e^{kt}$. This is not, however, the only dynamics that autocatalysts may follow. In fact, most non-enzymatic autocatalysts studied to date show a different behavior where their growth rate is sub-linear in their concentration and better described by $dA/dt = kA^n$ with $n < 1$~\cite{SemenovAC, von1986self, sievers_self-replication_1998, Sutherland1997, szathmary_sub-exponential_1989}. This corresponds to a slower, polynomial dynamics, of the form $A(t) \sim t^{1/(1-n)}$. The value $n\approx 1/2$ has most often been observed, leading to $A(t)\sim t^2 $, also known as parabolic growth~\cite{Kie2001}. In autocatalysis through template replication, this value is understood as arising from product inhibition, the common rebinding of a product to a template~\cite{Kie2001}. As far as growth and selection are concerned, however, the underlying mechanisms are not essential.


Instead, past works have stressed that the value of $n$ captures the most fundamental distinction, setting apart exponential ($n=1$) from sub-exponential ($n<1$) autocatalytic \obj~\cite{szathmary_sub-exponential_1989, Stadler1998, szathmary1991simple}. While exponential \obj~with different $k$ are mutually exclusive, sub-exponential ones generally coexist, with exclusion taking place only in particular limiting cases~\cite{Stadler1998, LL99, Kie2000, Sza2001}. As a consequence of this result, the field of experimental abiogenesis and the broader community engaged in developing non-enzymatic autocatalysts have concentrated their efforts on creating autocatalysts that can achieve exponential growth~\cite{Kie2001, SemenovAC, robertson_minimal_2000, Issac2002, Ouldrige2022, Otto2015, virgo_evolvable_2012, zeravcic_self-replicating_2014, dempster_self-replication_2015, zhuo_litters_2019}. However, the theoretical studies on which this conclusion is based have only considered autocatalytic \obj~with different growth parameters $k$ and the same exponent $n$, without exploring the possibility for $n$ to differ between competitors. Yet, $n$ is well-recognized to be, as much as $k$, an effective parameter that can vary between autocatalytic \obj~and can therefore be subject to selection. Here, we extend previous analyses to study the selection of competing autocatalytic \obj~with different exponents $n$, in addition to different parameters $k$. As we show, the results are counterintuitive and challenge the view that only exponential autocatalytic \obj~can be excluding: sub-exponential autocatalytic \obj~can exclude exponential ones, but not vice versa.

\section{Models}

The simplest setting to study autocatalysis under resource limitation is that of a continuous stirred-tank reactor, or chemostat~\cite{novick1950description}, where the resource needed for reproduction is introduced and removed at a constant rate, such that its concentration $R$ is coupled to the concentration $A$ of the autocatalytic \obj~by
\bea
\begin{aligned}
   \frac{dA}{dt}&=kRA^n - DA,\\
    \frac{dR}{dt}&=D(R_0-R)-kRA^n.
\end{aligned}
\eea
\noindent In these equations, $k$ denotes the \obj~replication rate constant, $n$ signifies its reaction order, and $D$ serves a dual purpose: it represents both a common dilution or decay rate for the autocatalytic \obj~and the resource, as well as the rate at which the resource is replenished from a reservoir with concentration $R_0$. In what follows, we introduce $\tau = Dt$ and $\kappa = k/D$ to have effectively $D=1$,

\bea
\begin{aligned} \label{Eq: lonely}
    \frac{dA}{d\tau}&= \kappa RA^n - A,\\
     \frac{dR}{d\tau}&=R_0-R-\kappa RA^n.
\end{aligned}
\eea

The minimal concentration of resource that allows an autocatalytic \obj~to grow ($dA/d\tau > 0 $) is $R = A^{1-n}/\kappa$. When growth is sub-exponential ($n<1$), $A$ therefore grows whenever it is small enough. \ys{More specifically, a stability analysis indicates that the boundary equilibrium $A=0$ is always unstable when $n<1$ ($\lim_{A \to 0} d^2A/d\tau^2 = +\infty$):} this implies that sub-exponential autocatalytic \obj~can never become extinct. When, instead, growth is exponential ($n=1$), survival solely depends on the concentration of resource and is possible only if $R > 1/\kappa$; \ys{correspondingly, the boundary equilibrium $A=0$ is unstable only if $R > 1/\kappa$ in this case.} This fundamental difference is the key to understanding why sub-exponential autocatalytic \obj~can coexist while exponential ones exclude each others but also, as we show below, why sub-exponential autocatalytic \obj~can exclude exponential ones but not conversely.

To study exclusion and coexistence \ys{of \obj~subject to a common limiting resource}, we extend the model to include two \obj~dependent on the same resource,
\bea\label{eq: Chemostat}
\begin{aligned}
    \frac{dA_1}{d\tau}&= \kappa_1RA_1^{n_1} - A_1,\\
    \frac{dA_2}{d\tau}&= \kappa_2RA_2^{n_2} - A_2,\\
    \frac{dR}{d\tau}&= R_0-R-\sum_{i=1}^2\kappa_i RA_i^{n_i}.
\end{aligned}
\eea
Here, we view the concentration $R_0$ of resource in the reservoir as an extrinsic or ``environmental'' parameter, but the parameters $\kappa_i$ and $n_i$ as parameters intrinsic to each autocatalytic \obj~$i$ and therefore potentially subject to selection.
Our point is to identify the intrinsic and extrinsic conditions that lead to the exclusion of one \obj~by another. To this end, we consider that a first, resident \obj~has reached a steady state and analyze whether a second, invading \obj, can grow when introduced in infinitesimal quantity in the background of the resident one~\cite{szathmary1991simple, LL2001, Sza2001}.

\section{Results}

When the two autocatalytic \obj~are exponential ($n_1=n_2=1$), the invading \obj~faces a concentration of resource $\bar R_1$ set by the resident autocatalysts with $\bar R_1=1/\k_1$ if $\k_1>1/R_0$ and $\bar R_1=R_0$ otherwise, in which case the resident autocatalysts do not survive by themselves. If $\k_2<\k_1$, the invading \obj~cannot grow, while if $\k_2>\k_1$ and $\k_2>1/R_0$ it grows to eventually exclude the resident \obj. This is the essence of the exclusion principle~\cite{szathmary_sub-exponential_1989}: two exponential autocatalytic \obj~cannot coexist if they depend on the same resource and have different replication rate constants.

If the resident \obj~is exponential ($n_1=1$) but the invading one is sub-exponential ($n_2<1$), however, the situation is different since the sub-exponential \obj~can always grow provided its concentration is low enough, i.e., provided $A_2<(R/\k_2)^{1/(1-n_2)}$. There is therefore no way for the resident  exponential autocatalytic \obj~to exclude an invading sub-exponential \obj. If, on the other hand, the resident \obj~is sub-exponential and the invading \obj~is exponential, two scenarios are possible. If $\kappa_1>1/\bar R_2$, where $\bar R_2$ is the steady state concentration of resource in presence of the sub-exponential \obj~alone, the exponential \obj~can invade and come to coexist with the sub-exponential one. If, however, $\kappa_1<1/\bar R_2$, the exponential \obj~cannot invade and is therefore excluded. 

\begin{figure}[t]
\centering
\includegraphics[width=.9\linewidth]{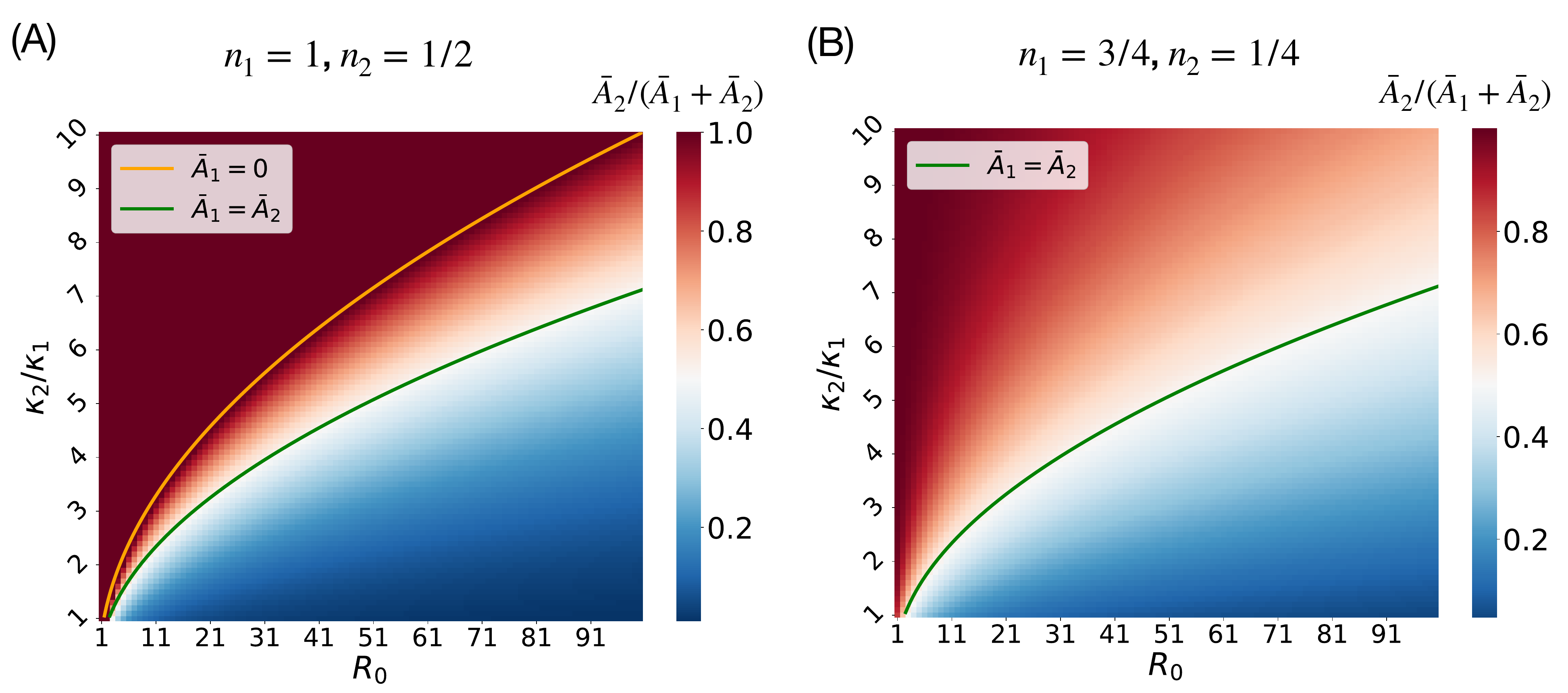}
\caption{\footnotesize Exclusion and coexistence in competitions of autocatalytic \obj~with different growth laws, i.e., different exponents $n$. {\bf A.} Mixture of a exponential and a sub-exponential autocatalytic \obj: $n_1=1$ and $n_2=1/2$. {\bf B.} Mixture of two sub-exponential \obj: $n_1=3/4$ and $n_2=1/4$. The graphs show the relative concentration at steady state of one of two \obj, $\bar{A}_2/(\bar{A}_2+\bar{A}_1)$, as a function of the total concentration of resource $R_0$ and of the ratio $\kappa_2/\kappa_1$, when $\kappa_1 = 1$. The steady-state concentrations are obtained by numerical integration of the kinetic equations, Eq.~\eqref{eq: Chemostat}. In A, the orange line indicates the total concentration of resource $R_0$ below which the exponential autocatalytic \obj~is excluded  ($\bar{A}_1 = 0$). The green lines indicate the total concentration $R_0$ for which the two \obj~are in same proportion, $\bar{A}_1 = \bar{A}_2$.} \label{fig: effective_model}
\end{figure}

Beyond this invasion analysis, the condition for exclusion of a exponential ($n_1=1$) autocatalytic \obj~by a sub-exponential ($n_2<1$) one is
\bea\label{eq: condition}
R_0 < 1/\kappa_1+(\kappa_2/\kappa_1)^\frac{1}{1-n_2}.
\eea
This includes, in particular, levels of resource $R_0$ where the exponential \obj~would survive by itself but is excluded by the sub-exponential \obj, when $1/\kappa_1<R_0<1/\kappa_1+(\kappa_2/\kappa_1)^\frac{1}{1-n_2}$. This is illustrated in Fig.~\ref{fig: effective_model}A with the case of an exponentially growing autocatalytic \obj~($n_1=1$) competing with a parabolically growing \obj~ ($n_2=1/2$).

Finally, when two sub-exponential \obj~($n_1<1$ and $n_2<1$) compete, they necessarily coexist. In the particular case where they have same exponent ($n_1=n_2=n$), their relative concentration at steady state only depends on their intrinsic parameters, $\bar{A}_1/\bar{A}_2 = (\kappa_1/\kappa_2)^{1/(1-n)}$. More generally, however, their relative concentration is controlled by the amount of resource $R_0$. In this case again, when $R_0$ is low, the most abundant one can be, somewhat counterintuitively, the one with lowest exponent $n$ (Fig.~\ref{fig: effective_model}B).

\ys{The analysis can be extended to investigate the conditions under which a sub-exponential autocatalytic \obj~($n<1$) can exclude a super-exponential one ($n>1$). The minimal concentration of resource that allows super-exponential \obj~to grow ($dA/d\tau > 0$) increases as their concentration decreases, $R=1/(A^{n-1}\kappa)$. Therefore, in a situation where the resident \obj~is sub-exponential, an autocatalytic \obj~introduced in infinitesimal quantity is less advantaged when it is super-exponential than when it is exponential. However, a resident \obj~requires a smaller concentration of resource when it is super-exponential than when it is exponential, and it is therefore more difficult to displace. The results of the competition of a super-exponential \obj~with another \obj~therefore depends strongly on the initial conditions. This has been referred to as the ``survival of the first'' in previous studies~\cite{Plasson}. Autocatalytic \obj~growing super-exponentially have, however, not been demonstrated experimentally so far, even though theoretical models that include hypercycles could result in it~\cite{eigen_selforganization_1971,eigen_principle_1977,Eigen2013, szathmary_propagation_2013}.}

\section{Discussion}

The primary aim of this paper is to highlight that a sub-exponential autocatalytic \obj~can exclude an exponential autocatalytic \obj~and, more generally, that a sub-exponential \obj~can dominate one of higher replication order, i.e., following a growth law $dA/dt= kA^n$ with a larger exponent $n$. This selection for autocatalytic \obj~of lower order occurs because autocatalytic dynamics depends not only on the growth order $n$, but also on the replication rate constant $k$: a \obj~with a lower reaction order can outcompete another with a higher reaction order if it has a greater replication rate constant.

Our results rely on the phenomenological equation $dA/dt= kA^n$, which is widely employed to model the competitive dynamics of autocatalysts dependent on a common limiting resource~\cite{szathmary_sub-exponential_1989, szathmary1991simple, LL99, Kie2000, Sza2001, SemenovAC}. In practice, the parameters $k$ and $n$ may be either inferred from experimental data~\cite{von1986self, sievers_self-replication_1998, Issac2002, Otto2015, zhang_engineering_2007}, or derived from mechanistic models~\cite{Kie2001}. Derivations from mechanistic models bring an important nuance by showing that sub-exponential growth typically arises as an approximation of a more general relationship $dA/dt = f(A)$, where $f(A) \sim A^n$ with $n<1$ for sufficiently high concentrations of $A$~\cite{Kie2001, Sza2001, Stadler1998}. At low concentration, however, this sub-exponential growth typically turns into an exponential growth, since $dA/dt \approx f'(0)A$ for small values of $A$. The presence of these two regimes is well understood when sub-exponential growth stems from product inhibition~\cite{Kie2001, sievers_self-replication_1998}, which is necessarily negligible at low autocatalyst concentration.  \ys{This dependence of the growth rate on the concentration of autocatalyst is generic and, in the absence of mechanistic details, autocatalytic growth can typically be captured through logistic or similarly shaped functions, for which the growth of an autocatalyst is exponential at low concentration and decreases at higher concentration~\cite{Schuster, Bentea}. In any case, this implies that a} sub-exponential autocatalytic \obj~following a growth law with these two regimes becomes extinct below a certain concentration, which opens the possibility for an exponential one to exclude it. This does not affect, however, the possibility for the same sub-exponential \obj~to exclude an exponential \obj, as we have noted.

Current research on autocatalysis is geared towards the experimental design of exponential autocatalytic \obj, motivated by the desire to observe exclusion, which is viewed as a key step towards achieving evolution by natural selection~\cite{Sza97, LL97, Stadler1998, Otto2015}. The exclusion principle, first formulated by Gause in an ecological context~\cite{Gause}, states that among two exponential \obj, the one with the higher replication rate will outcompete the other, irrespective of the magnitude of the difference between the rates~\cite{szathmary_sub-exponential_1989}. Yet, our findings indicate that sub-exponential autocatalytic \obj~can outperform their exponential \ys{and even super-exponential} counterparts, and can therefore also cause selection by exclusion.
As such, sub-exponential autocatalytic \obj~set constraints on the evolutionary emergence of the first exponential ones. The issue takes on particular importance when we recognize the interrelated nature of the parameters \(k\) and \(n\) arising from mechanistic models. These parameters are indeed influenced by shared physical factors like reaction volume and autocatalyst-substrate interaction strength, resulting in a trade-off between them. 
Given that we have shown that \(k\) can be the primary determinant of autocatalyst dominance, this type of physical correlation further calls into question an exclusive emphasis on exponential autocatalysts.

\acknowledgments

We thank Z. Zeravcic for comments on an earlier draft. This work was supported by ANR-22-CE06-0037. Declarations of interest: none


\begin{thebibliography}{0}
\expandafter\ifx\csname natexlab\endcsname\relax\def\natexlab#1{#1}\fi
\expandafter\ifx\csname bibnamefont\endcsname\relax
  \def\bibnamefont#1{#1}\fi
\expandafter\ifx\csname bibfnamefont\endcsname\relax
  \def\bibfnamefont#1{#1}\fi
\expandafter\ifx\csname citenamefont\endcsname\relax
  \def\citenamefont#1{#1}\fi
\expandafter\ifx\csname url\endcsname\relax
  \def\url#1{\texttt{#1}}\fi
\expandafter\ifx\csname urlprefix\endcsname\relax\def\urlprefix{URL }\fi
\providecommand{\bibinfo}[2]{#2}
\providecommand{\eprint}[2][]{\url{#2}}

\end{thebibliography}


\begin{thebibliography}{10}

\bibitem{SemenovAC}
Anton I. Hanopolskyi, Viktoryia A. Smaliak, Alexander I. Novichkov, and Sergey
  N. Semenov.
\newblock Autocatalysis: Kinetics, mechanisms and design.
\newblock {\em ChemSystemsChem}, 3(1), e2000026, 2020.

\bibitem{von1986self}
G{\"u}nter von Kiedrowski.
\newblock A self-replicating hexadeoxynucleotide.
\newblock {\em Angewandte Chemie International Edition in English},
  25(10):932--935, 1986.

\bibitem{sievers_self-replication_1998}
Dirk Sievers and G{\"u}nter Von~Kiedrowski.
\newblock Self-{Replication} of {Hexadeoxynucleotide} {Analogues}:
  {Autocatalysis} versus {Cross}-{Catalysis}.
\newblock {\em Chemistry - A European Journal}, 4(4):629--641, 1998.

\bibitem{Sutherland1997}
Bing Wang and Ian~O. Sutherland.
\newblock Self-replication in a diels–alder reaction.
\newblock {\em Chemical Communications}, 16: 1495-1496., 1997.

\bibitem{szathmary_sub-exponential_1989}
E{\"o}rs Szathm{\'a}ry and Irina Gladkih.
\newblock Sub-exponential growth and coexistence of non-enzymatically
  replicating templates.
\newblock {\em Journal of Theoretical Biology}, 138(1):55--58, 1989.

\bibitem{Kie2001}
Gunter von Kiedrowski.
\newblock Minimal replicator theory i: Parabolic versus exponential growth.
\newblock {\em Bioorganic Chemistry Frontiers, vol 3. Spring}, 1993.

\bibitem{Stadler1998}
Peter~R. Wills, Stuart~A. Kauffman, Barbel M.~R. Stadler, and Peter~F. Stadler.
\newblock Selection dynamics in autocatalytic systems: Templates replicating
  through binary ligation.
\newblock {\em Bulletin of mathematical biology}, 118(60):1073--1098, 1998.

\bibitem{szathmary1991simple}
E{\"o}rs Szathm{\'a}ry.
\newblock Simple growth laws and selection consequences.
\newblock {\em Trends in Ecology \& Evolution}, 6(11):366--370, 1991.

\bibitem{LL99}
Shneior Lifson and Hanna Lifson.
\newblock A model of prebiotic replication: Survival of the fittest versus
  extinction of the unfittest.
\newblock {\em Journal of theoretical biology}, 199(4):425--433, 1999.

\bibitem{Kie2000}
G.~von Kiedrowski and E{\"o}rs Szathm{\'a}ry.
\newblock Selection versus coexistence of parabolic replicators spreading on
  surfaces.
\newblock {\em Selection}, 1(3):173--180, 2001.

\bibitem{Sza2001}
Istvan Scheuring and E{\"o}rs Szathm{\'a}ry.
\newblock Survival of replicators with parabolic growth tendency and
  exponential decay.
\newblock {\em Journal of theoretical biology}, 212(1):99--105, 2001.

\bibitem{robertson_minimal_2000}
Andrew Robertson, Andrew J.~Sinclair, and Douglas Philp.
\newblock Minimal self-replicating systems.
\newblock {\em Chemical Society Reviews}, 29(2):141--152, 2000.
\newblock Publisher: Royal Society of Chemistry.

\bibitem{Issac2002}
Roy Issac and Jean Chmielewski.
\newblock Approaching exponential growth with a self-replicating peptide.
\newblock {\em Journal of the American Chemical Society},  124(24), 6808-6809, 2002.

\bibitem{Ouldrige2022}
Jordan Juritz, Jenny~M. Poulton, and Thomas~E. Ouldridge.
\newblock Minimal mechanism for cyclic templating of length-controlled
  copolymers under isothermal conditions.
\newblock {\em J. Chem. Phys.}, 156(7), 2022.

\bibitem{Otto2015}
Mathieu Colomb-Delsuc, Elio Mattia, Jan~W. Sadownik, and Sijbren Otto.
\newblock Exponential self-replication enabled through a fibre
  elongation/breakage mechanism.
\newblock {\em Nature Communication}, 6(7427), 2015.

\bibitem{virgo_evolvable_2012}
Nathaniel Virgo, Chrisantha Fernando, Bill Bigge, and Phil Husbands.
\newblock Evolvable {Physical} {Self}-{Replicators}.
\newblock {\em Artificial Life}, 18(2):129--142, 2012.

\bibitem{zeravcic_self-replicating_2014}
Zorana Zeravcic and Michael~P. Brenner.
\newblock Self-replicating colloidal clusters.
\newblock {\em Proceedings of the National Academy of Sciences},
  111(5):1748--1753, 2014.

\bibitem{dempster_self-replication_2015}
Joshua~M. Dempster, Rui Zhang, and Monica Olvera De La~Cruz.
\newblock Self-replication with magnetic dipolar colloids.
\newblock {\em Physical Review E}, 92(4):042305, 2015.

\bibitem{zhuo_litters_2019}
Rebecca Zhuo, Feng Zhou, Xiaojin He, Ruojie Sha, Nadrian~C. Seeman, and Paul~M.
  Chaikin.
\newblock Litters of self-replicating origami cross-tiles.
\newblock {\em Proceedings of the National Academy of Sciences},
  116(6):1952--1957, 2019.

\bibitem{novick1950description}
Aaron Novick and Leo Szilard.
\newblock Description of the chemostat.
\newblock {\em Science}, 112(2920):715--716, 1950.

\bibitem{LL2001}
Shneior Lifson and Hanna Lifson.
\newblock Coexistence and darwinian selection among replicators: Response to
  the preceding paper by scheuring and szathmary.
\newblock {\em Journal of molecular evolution}, 212(1):107--109, 1997.

\bibitem{Plasson}
Raphael Plasson, Axel Brandenburg, Ludovic Jullien, Hugues Bersini.
\newblock Autocatalysis: At the Root of Self-Replication
\newblock {\em Artificial Life}, 17(3): 219–236, 2011.

\bibitem{eigen_selforganization_1971}
Manfred Eigen.
\newblock Selforganization of matter and the evolution of biological
  macromolecules.
\newblock {\em Naturwissenschaften volume}, 58:465--523, 1971.

\bibitem{eigen_principle_1977}
Manfred Eigen and Peter Schuster.
\newblock A principle of natural self-organization.
\newblock {\em Naturwissenschaften}, 64(11):541--565, 1977.

\bibitem{Eigen2013}
Manfred Eigen.
\newblock {\em From Strange Simplicity to Complex Familiarity}.
\newblock Oxford University Press, 2013.

\bibitem{szathmary_propagation_2013}
E{\"o}rs Szathm{\'a}ry.
\newblock On the propagation of a conceptual error concerning hypercycles and
  cooperation.
\newblock {\em Journal of Systems Chemistry}, 4(1):1, 2013.

\bibitem{zhang_engineering_2007}
David~Yu Zhang, Andrew~J Turberfield, Bernard Yurke, and Erik Winfree.
\newblock Engineering {Entropy}-{Driven} {Reactions} and {Networks} {Catalyzed}
  by {DNA}.
\newblock 318, 2007.

\bibitem{Schuster}
Peter Schuster.
\newblock What is special about autocatalysis?
\newblock {\em Monatsh Chem}, 150:763–775, 2019.

\bibitem{Bentea}
Lucian Bentea, Murielle A. Watzky, Richard G. Finke.
\newblock Sigmoidal Nucleation and Growth Curves Across Nature Fit by the Finke-Watzky Model of Slow Continuous Nucleation and Autocatalytic Growth: Explicit Formulas for the Lag and Growth Times Plus Other Key Insights.
\newblock {\em J. Phys. Chem. C}, 121(9):5302–5312, 2017.

\bibitem{Sza97}
E{\"o}rs Szathm{\'a}ry and John Maynard~Smith.
\newblock From replicators to reproducers: the first major transitions leading
  to life.
\newblock {\em Journal of theoretical biology}, 187(4):555--571, 1997.

\bibitem{LL97}
Shneior Lifson.
\newblock On the crucial stages in the origin of animate matter.
\newblock {\em Journal of theoretical biology}, 44(1):1--8, 2001.

\bibitem{Gause}
J.F Gause.
\newblock {\em The struggle for existence}.
\newblock The Williams \& Wilkins company, 1934.

\end{thebibliography}
\end{document}